\documentclass[aps,pra,twocolumn,superscriptaddress,showpacs]{revtex4}
\usepackage{amsmath}
\usepackage{epsfig}
\usepackage{cancel}
\usepackage{hyperref}
\usepackage[usenames]{color}
\begin{document}
\title{Heteronuclear molecules in an optical dipole trap}

\author{J. J. Zirbel}
 \email{zirbel@jilau1.colorado.edu}
\affiliation{JILA, Quantum Physics Division, National Institute of
Standards and Technology and Department of Physics, University of
Colorado, Boulder, CO 80309-0440, USA}
\author{K.-K. Ni}
\affiliation{JILA, Quantum Physics Division, National Institute of
Standards and Technology and Department of Physics, University of
Colorado, Boulder, CO 80309-0440, USA}
\author{S. Ospelkaus}
\affiliation{JILA, Quantum Physics Division, National Institute of
Standards and Technology and Department of Physics, University of
Colorado, Boulder, CO 80309-0440, USA}
\author{T. L. Nicholson}
\affiliation{JILA, Quantum Physics Division, National Institute of
Standards and Technology and Department of Physics, University of
Colorado, Boulder, CO 80309-0440, USA}
\author{M. L. Olsen}
\affiliation{JILA, Quantum Physics Division, National Institute of
Standards and Technology and Department of Physics, University of
Colorado, Boulder, CO 80309-0440, USA}
\author{P. S. Julienne}
\affiliation{Joint Quantum Institute, National Institute of Standards and Technology and University of Maryland, Gaithersburg, Maryland 20899-8423, USA}
\author{C. E. Wieman}
\affiliation{JILA, Quantum Physics Division, National Institute of
Standards and Technology and Department of Physics, University of
Colorado, Boulder, CO 80309-0440, USA}
\author{J. Ye}
\affiliation{JILA, Quantum Physics Division, National Institute of
Standards and Technology and Department of Physics, University of
Colorado, Boulder, CO 80309-0440, USA}
\author{D. S. Jin}
\affiliation{JILA, Quantum Physics Division, National Institute of
Standards and Technology and Department of Physics, University of
Colorado, Boulder, CO 80309-0440, USA}

\date{\today}

\begin{abstract}
We report on the creation and characterization of heteronuclear
$^{40}$K$^{87}$Rb Feshbach molecules in an optical dipole
trap. Starting from an ultracold
gas mixture of $^{40}$K and $^{87}$Rb atoms, we create as many as 25,000 molecules at 300\,nK by rf association. Optimizing the association process, we achieve a conversion
efficiency of 25\%. We measure the temperature dependence of the rf association process and find good agreement with a phenomenological model that has
previously been applied to Feshbach molecule creation by slow
magnetic-field sweeps.  We also present a  measurement of the
binding energy of the heteronuclear molecules in the
vicinity of the Feshbach resonance and provide evidence for Feshbach
molecules as deeply bound as 26\,MHz.
\end{abstract}


\maketitle

\section{Introduction}

Magnetic-field tunable Feshbach resonances in ultracold atomic gases
provide a unique tool for the efficient conversion of ultracold
atoms into weakly bound diatomic molecules \cite{Kohler2006}.  Most
studies of Feshbach molecule formation have been performed using
single-species atomic gases. However, heteronuclear molecules have
bright prospects as quantum gases with long-range anisotropic
interactions \cite{Santos2000}, as quantum bits for novel quantum computation schemes \cite{Demille2002}, and as sensitive probes for precision measurements \cite{Sandars1967,Kozlov1995, Hudson2006}.
These applications make use of the fact that a heteronuclear molecule
can have an electric dipole moment.

Production of ultracold, polar molecules presents significant
technical hurdles because of the rich internal molecular structure.
Although there are other methods of creating ultracold polar molecules \cite{Bethlem1999,Weinstein1998,Rangwala2003,Kerman2004a,Sawyer2007, Kleinert2007}, one strategy is to start from a nearly quantum degenerate atom gas and associate the atoms into molecules using time-dependent magnetic fields in
the vicinity of a Feshbach resonance \cite{Donley2002, Regal2003, Jochim2003a, Xu2003, Herbig2003, Inouye2004,Papp2006, Thalhammer2006, Deh2008, Wille2008}. These Feshbach
molecules have the advantage of starting in a single
internal state at ultralow temperature. However, they are
extremely weakly bound and therefore do not have a significant
electric dipole moment.  It may be possible to employ coherent optical transfer schemes \cite{Pe'er2007} to transfer these molecules into a more deeply bound state and thus create an ultracold sample of polar molecules \cite{Kotochigova2003}.

Creation of $^{40}$K$^{87}$Rb Feshbach molecules was recently
reported by Ospelkaus and co-workers \cite{Ospelkaus2006}.  In that
work, the atoms were confined in a 3D optical lattice potential with
precisely one atom pair per site.  Confinement in an optical
lattice has two advantages.  First, the tight confinement of a pair
of atoms in each lattice site meant that the initial motional state of each
atom pair was the ground state of the trapping potential at that
site. Magnetoassociation can then proceed by driving a transition
between two well defined quantum states. Second, confinement of the
atoms and molecules in the optical lattice protected the Feshbach
molecules from collisions.  This protection is important because the
extremely weakly bound molecules can suffer heating and loss through
inelastic collisions that cause vibrational quenching.

In this article, we report on the creation of fermionic
heteronuclear $^{40}$K$^{87}$Rb Feshbach molecules in a far-detuned
optical dipole trap.  In this relatively weak trap, the atoms start
in a near-continuum of motional states, and furthermore, atoms and
molecules are free to collide. Surprisingly, without using
confinement in an optical lattice, we find similar overall
efficiency in the radio frequency (rf) association of ultracold heteronuclear molecules and
can create 25,000 molecules at a temperature of about
300 nK.   We study the association process and present a measurement of the magnetic-field dependent binding energy of the Feshbach
molecules. Compared to previous results \cite{Ospelkaus2006}, we extend
our measurements to magnetic fields much farther from the Feshbach
resonance and also discuss basic properties of the molecules such as
their size and quantum state composition. These results, as well as our recent study of the collisional stability of these molecules \cite{Zirbel2007a}, can provide a starting point for future experiments that will produce polar molecules by driving Feshbach molecules to more deeply bound states.

The paper is organized in the following way: In Section II, we
discuss the preparation of a near quantum degenerate Fermi-Bose mixture
of $^{40}$K and $^{87}$Rb atoms.  In Section III, we present studies of
molecule production using rf magnetoassociation.  We look at the
temperature dependence of the association process and compare our
results to a simple phenomenological model. In Section IV we discuss measurements of the properties of the molecules such as their thermal energy and trapping frequency.  In Sections V and VI, we present measurements of the magnetic-field dependent binding energy of the heteronuclear molecules and compare our experimental results to a coupled-channel calculation. Knowledge of the binding energy allows us to predict basic properties of the molecules such as their size and their open and closed channel fractions.

\section{Production of an ultracold Fermi-Bose atom mixture}

Efficient association of ultracold atoms into molecules near a magnetic-field
Feshbach resonance requires an ultracold atom gas near quantum
degeneracy \cite{Hodby2005}. We use a two-species vapor-cell
magneto-optical trap (MOT) to initially cool and trap fermionic
$^{40}$K atoms and bosonic $^{87}$Rb atoms, similar to ref.
\cite{Goldwin2002}. After loading the MOT, the atoms are transferred
into a movable quadrupole magnetic trap for delivery to a
differentially pumped, ultra-low pressure section of the vacuum
chamber \cite{Lewandowski2003}. Here, the atoms are transferred into
a Ioffe-Pritchard type magnetic trap for evaporative cooling. This
trap provides harmonic confinement with axial and radial trap
frequencies for Rb of 20 Hz and 156 Hz, respectively.  Forced microwave
evaporation, which drives a hyperfine spin-flip transition, is used to
directly cool the Rb gas, which in turn sympathetically cools the K
gas.

Controlling the atoms' spin states is important for collisional
stability of the two species mixture.  Before loading the quadrupole
magnetic trap, the atoms are optically pumped into the stretched
states, which experience the largest trapping force. These are the
$|2,2\rangle$ and $|9/2,9/2\rangle$ states for Rb and K,
respectively, where $|f,m_{f}\rangle$ denotes the quantum state of the atom having total atomic spin,
$f$, and projection along the local magnetic-field direction,
$m_{f}$. To remove any Rb atoms remaining in other spin
states, the atoms are loaded into the quadrupole magnetic trap with a gradient of 20 G/cm along the vertical
direction. Rb atoms in states other than the stretched states are
not supported against gravity and are thus filtered out. We also
find that our evaporation efficiency for K atoms improves
significantly if we continuously remove any Rb atoms in the
$|2,1\rangle$ state. For this, we apply a fixed microwave frequency
tuned to the $|2,1\rangle \rightarrow |1,0\rangle$ transition
frequency for atoms at the trap center.

We start with $10^{9}$ Rb atoms and $10^{7}$ K atoms in the magnetic
quadrupole trap, and after evaporation in the Ioffe-Pritchard magnetic
trap we have $5\times 10^{6}$ Rb and $1\times 10^{6}$ K
atoms at 3 $\mu$K.   These atoms are loaded into a far-off-resonance optical
dipole trap with a 50 $\mu$m waist derived from a single-frequency fiber laser operating at a wavelength of 1064 nm.  Once the optical dipole trap is loaded and the magnetic trap is turned off, we use rf adiabatic rapid passage to transfer the atoms into the
$|1,1\rangle$ and $|9/2,-7/2\rangle$ states of Rb and K,
respectively. The gases are then simultaneously evaporated at a
field of B=540 G by lowering the optical power in the trapping beam until
$3\times 10^{5}$ Rb at $T/T_{c} \sim 1.1$ and $1\times 10^{5}$ K at
$T/T_{F} \sim 0.6$ remain. $T_{c}$ is the critical temperature for the onset of Bose-Einstein condensation (BEC) and $T_{F}$ is the Fermi temperature. At the end of the optical trap evaporation we adiabatically increase the optical trap power until the measured radial trapping frequencies are 211 Hz and 136 Hz for K and Rb, respectively and the temperature is 150 nK.  We calculate that gravitational sag causes the in-trap center positions of the Rb and K to be separated by 8\,$\mu$m. The calculated trap depths are 1.1\,$\mu$K for Rb and 2.8\,$\mu$K for K.

We extract the number and temperature of the atoms from time-of-flight, resonant absorption images. The gas is suddenly released from the trap and allowed to expand before imaging. Using a resonant light pulse of the appropriate frequency, we selectively image K or Rb atoms in a particular spin state.

\section{Feshbach molecule production by rf association}

To create molecules, we utilize a Feshbach resonance between Rb
$|1,1\rangle$ and K $|9/2,-9/2\rangle$ atoms at $B_{0}=$546.76 G
\cite{Inouye2004}.  Similar to previous work using this resonance
\cite{Ospelkaus2006}, we start with atoms at a magnetic field near
the Feshbach resonance and apply a transverse rf magnetic field to
convert atom pairs to molecules.  However, unlike the work of ref.
\cite{Ospelkaus2006}, in our experiment the atom pairs are not
confined in an optical lattice.  Therefore, the atoms are free to
collide.

\begin{figure}
  \centering
  \includegraphics[width=\columnwidth]{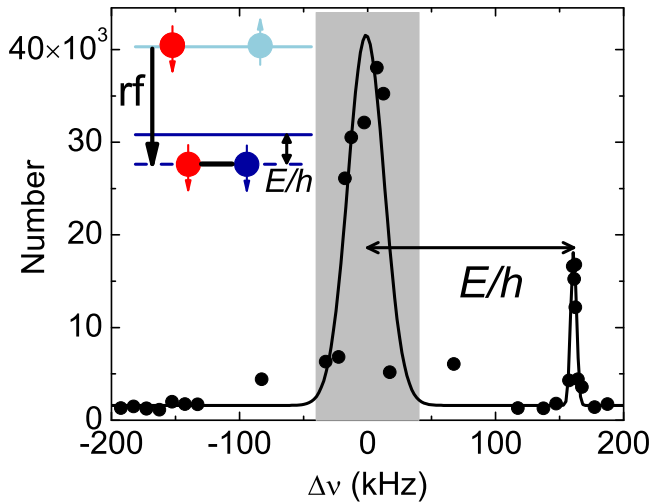}
  \caption{Rf association spectrum of heteronuclear Feshbach molecules at $B=546.04$\,G.  The molecules are created starting from an ultracold gas of $1.7\times 10 ^{5}$ $^{87}$Rb atoms
  at $T/T_c=1$ in thermal equilibrium with $8\times10^4$ $^{40}$K atoms.
  The plot shows the detected atom/molecule number as a function of the detuning
  $\Delta\nu$ of the rf from the atomic resonance
  frequency.  We observe a large atomic peak at zero detuning and a
  molecular peak at $E/h= 161$\,kHz due to the association
  of atom pairs into the Feshbach molecule state.  In this measurement the rf
  pulse duration was 600 $\mu$s, which is 50 times longer than an
atomic $\pi$-pulse.  The grey band indicates the expected envelope
full width at half maximum for the power broadened atomic transition.  The inset shows a simplified energy diagram for magnetic fields below the Feshbach resonance.  The upper solid line corresponds to Rb $|1,1\rangle$ and K $ |9/2,-7/2\rangle$, while the lower solid line corresponds to Rb $|1,1\rangle$ and K $|9/2,-9/2\rangle$.  The molecule state is represented by the dashed line. Depending on the detuning, the rf is capable of driving an atomic hyperfine changing transition or magnetoassociation.
 }
\label{fig:rf_association_spectrum}
\end{figure}

In Fig. \ref{fig:rf_association_spectrum} we show a typical rf
association spectrum for a magnetic field tuned 0.7 G below the
Feshbach resonance.  The Rb and K atoms are initially prepared in
the $|1,1\rangle$ and $|9/2,-7/2\rangle$ states, respectively, and
we apply an rf pulse with a frequency tuned near the atomic K Zeeman
transition, $|9/2,-7/2\rangle\!\! \rightarrow \!\!|9/2,-9/2\rangle$.
For these data we apply a gaussian rf pulse with a 600 $\mu$s $1/e^{2}$ full-width and an rf strength that can drive a $12\,\mu$s $\pi$-pulse on the atomic K
Zeeman transition. The effect of the rf pulse is detected by counting the number of K atoms transferred to the previously unoccupied $|9/2,-9/2\rangle$ state.  For this purpose, we use spin-state selective absorption imaging of the cloud in the optical trap at $B=546$\,G.  We observe two peaks in the rf spectrum: one at the frequency of the atomic K Zeeman transition and one shifted to higher frequency.  The higher frequency peak
corresponds to a stimulated transition from a pair of atoms
into the Feshbach molecule state.  The shift of this spectral
feature from the atomic transition reveals the binding energy of
the molecule.  In comparing the widths of the two features it is important to note that the atom transition is power broadened by the rf pulse, which has a duration 50 times longer than the measured time for an atomic $\pi$-pulse.  The smaller width of the molecule feature indicates that the coupling strength is much weaker for molecule association than for the atomic Zeeman transition.  The full width at half maximum of the molecule feature is 7 kHz. This is consistent with the thermal distribution of the relative energies for atom pairs.

We are able to easily observe the molecular rf association feature because the Feshbach molecules can absorb the imaging light that is resonant with the K atom cycling transition at high magnetic field.  Presumably, the first photon dissociates the weakly bound molecule and subsequent photons scatter off the resulting K atom.  Compared to the case of homonuclear molecules, this direct imaging is possible for a larger range of Feshbach molecule binding energies.  This is because the energy separation between the excited and ground electronic potentials varies relatively slowly with internuclear distance, R, since the heteronuclear excited and ground electronic potentials both vary as $1/R^{6}$.  In contrast, for the homonuclear case, the excited electronic potential varies as $1/R^3$ and the ground potential varies as $1/R^6$.

\begin{figure}
  \centering
  \includegraphics[width=\columnwidth]{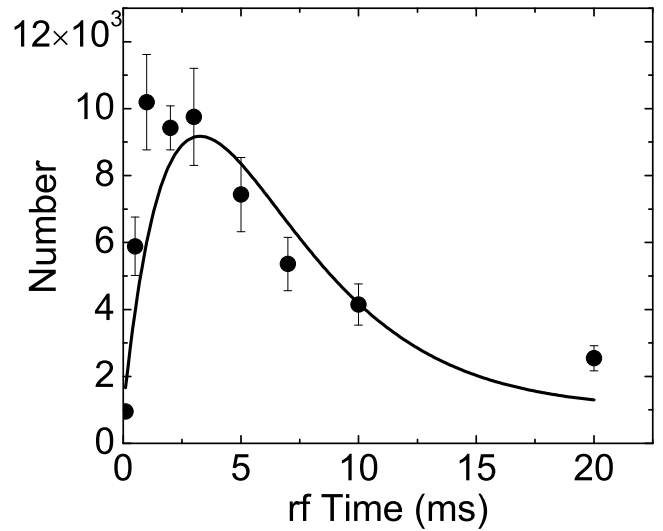}
  \caption{Time dependence of the rf association process at B\,=\,546.17\,G, where the measured molecule binding energy is $E/h=96$\,kHz. The plot
  shows the measured number of molecules created
  as a function of the $1/e^2$ gaussian rf pulse length. The solid line is a simple fit used to estimate the impact of the observed decay on the maximum number of molecules created.} \label{fig:time_dependence_rf}
\end{figure}

Because the atoms are free to move and collide in the trap, it is
interesting to look at the time dependence of the rf association.
The number of molecules observed as a function of the rf pulse duration is shown in Fig. \ref{fig:time_dependence_rf}. The gaussian $1/e^2$ width of the pulse is plotted on the horizontal axis.  For these data, the peak rf strength is fixed and only the duration of the rf pulse is changed.  The number of
molecules increases rapidly for the first few hundred microseconds
and then slowly decays.  We do not observe Rabi oscillations.  This is not surprising since we are driving transitions from a near continuum of free atom states to the bound molecule state. The observed decay results from atom-molecule collisions \cite{Zirbel2007a}.  To estimate the effect of this decay on the maximum number of molecules created we fit the data to the product of a linear growth and an exponential decay (curve in Fig. \ref{fig:time_dependence_rf}).  We find that the decays causes a 30\% reduction in the maximum number of molecules created at the peak of the curve in Fig. \ref{fig:time_dependence_rf}.  Since the maximum molecule creation in Fig. \ref{fig:time_dependence_rf} corresponds to approximately 10\% of the total number of K atoms, this raises the question of what limits the maximum conversion fraction?

\begin{figure}
  \centering
  \includegraphics[width=\columnwidth]{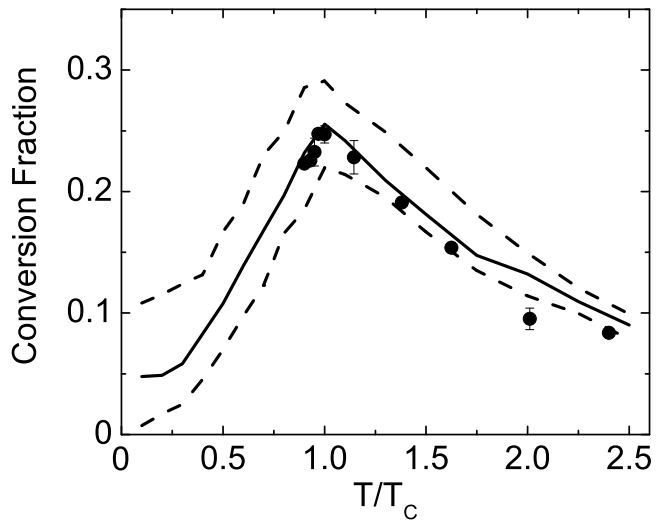}
  \caption{Temperature dependence of Feshbach molecule creation by rf association. Plotted is
  the molecule conversion fraction $N_{\mathrm{m}}/ N_{\mathrm{K}}$
  as a function of $T/T_{c}$ for the Rb atoms. The solid circles ($\bullet$) are experimental data and the curves are results of a Monte Carlo calculation. The dashed curves represent the uncertainty in the conversion fraction taking into account the uncertainty in the trap frequencies and atom numbers.  The molecules are created
  starting from an atom gas mixture with
  $N_{\mathrm{Rb}}=3\times 10^5$ and $N_{\mathrm{K}}=1\times10^5$. We observe
   maximum conversion efficiency of $25\%$ at $T/T_c=1$.}
\label{fig:rf_conversion_fraction}
\end{figure}

Previous studies of molecule creation by slow magnetic-field sweeps
across a Feshbach resonance have shown that the maximum conversion
fraction depends on the phase space density of the atoms \cite{Hodby2005}. Not surprisingly, for a two-species atom gas mixture, the
molecule conversion efficiency also depends on the spatial overlap
of the two atom clouds \cite{Papp2006}. We have measured the molecule conversion fraction as a function of temperature for Feshbach molecules produced using rf association at B=546.17\,G. For this measurement, the gas
mixture was initially cooled to a temperature below the critical
temperature, $T_{c}$, for the onset of Bose-Einstein condensation of
the Rb gas. The temperature is then varied simply by waiting a variable
amount of time in a sufficiently deep trap and letting the gas
slowly heat up before creating Feshbach molecules. With
this technique, the number of Rb and K atoms is kept nearly
constant.  To create Feshbach molecules we use rf adiabatic rapid passage and sweep the rf frequency from 80.132\,MHz to 80.142\,MHz in 2\,ms. The results are shown in Fig. \ref{fig:rf_conversion_fraction}.

We find that the maximum conversion fraction occurs at $T/T_c\approx1$.  Here,  $25\%$ of the K atoms are converted to molecules; this
corresponds to 25,000 molecules.  It is interesting to compare the
measured conversion fraction for rf association to the prediction of
a phenomenological model introduced in ref. \cite{Hodby2005}.  This
model, using experimentally determined parameters, has been
successful in predicting the molecule conversion efficiency for
Feshbach molecules created by slow magnetic-field sweeps \cite{Hodby2005, Papp2006}, but has not been compared to data for rf association.
Following the procedure outlined in \cite{Hodby2005} we calculate the
predicted maximum conversion efficiency using realistic trap
potentials and distributions of our atomic clouds.  The curves in Fig. \ref{fig:rf_conversion_fraction} show the
results of the calculation, multiplied by 0.70 to estimate the effect of the molecule loss seen in Fig.
\ref{fig:time_dependence_rf}. The dashed curves represent the uncertainty in the conversion fraction due to the uncertainty in the trap frequencies and atom numbers. The agreement between the model for association by slow magnetic-field sweeps and the data for rf association is quite good.  The predicted dependence on
temperature agrees with our experimental results for rf association.
The peak in the molecule conversion fraction as a function of
$T/T_c$ can be understood as a balance between two factors: phase-space density and the spatial overlap of the two clouds.  As $T/T_c$ decreases, the phase-space density increases but, below $T/T_c=1$, the onset of Rb BEC reduces the spatial overlap of the Rb cloud with the K cloud.  The effect of gravitational sag also reduces the spatial overlap of the two clouds and at $T/T_c=1$ the model predicts a 30\% reduction in molecule conversion efficiency when compared to the case with no gravitational sag.

\section{Ultracold, trapped molecules}

\begin{figure}
  \centering
  \includegraphics[width=\columnwidth]{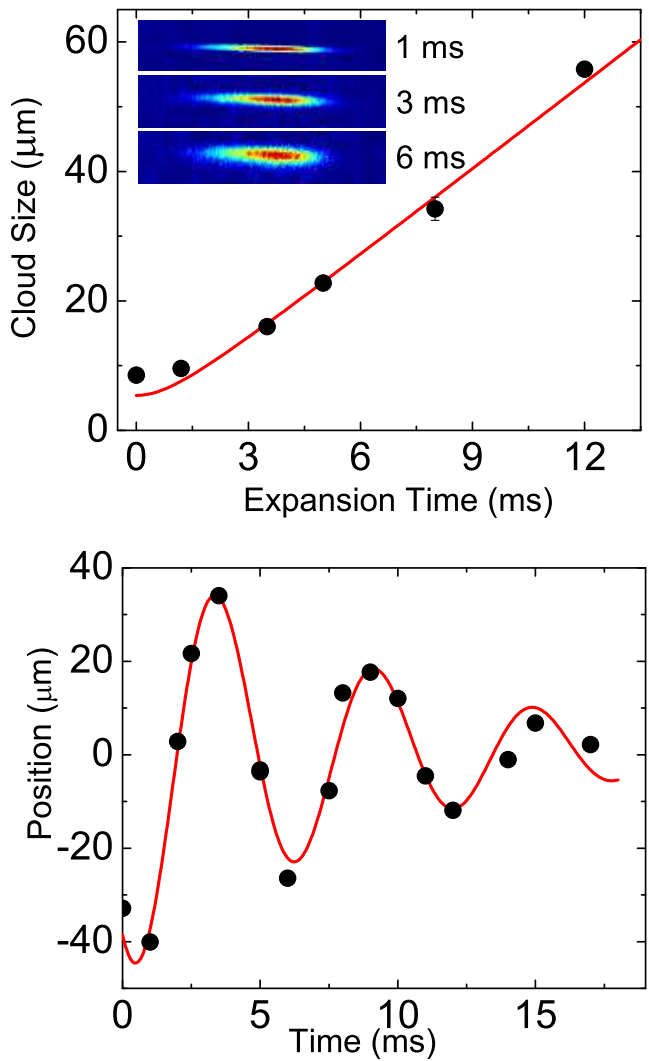}
  \caption{ (Top inset) Images of the molecule cloud after release from
  the optical trap. (Top) Radial cloud size for the molecules after
  release from the optical trap. (Bottom) Slosh of the molecular
  cloud in the optical trap seen in images of the expanded gas. The expansion time was 3 ms. The observed oscillation in the molecule cloud's position demonstrates
  that the molecules are indeed trapped.  The strong damping seen here is
  most likely due to anharmonicity of the optical trap.  } \label{fig:expansion}
\end{figure}

A feature of using rf association to create heteronuclear Feshbach
molecules is that we can selectively image the Feshbach molecules
using light resonant with the appropriate K atom spin-state. The
upper part of Fig. \ref{fig:expansion} shows the radial root-mean-squared size, $\sigma_{RMS}$, of the
molecular gas as a function of expansion time after the optical trap
is abruptly turned off. We fit the size to $\sigma_{RMS}(t)=(1/\omega_{r})\sqrt{k_{B} T/m_{\rm{KRb}}}\sqrt{1+\omega_{r}^{2}t^2}$ and obtain an expansion
energy of $T=310\pm20$\,nK. Here, $T$ is the only fit parameter, $k_{B}$ is Boltzmann's constant, $m_{\rm{KRb}}$ is the mass of the KRb molecule, and $\omega_{r}$ is the radial angular trapping frequency for the molecule.

To confirm that the KRb molecules are trapped,
we can look for sloshing of the molecule cloud after a perturbation
to the trap potential.  The lower part of Fig. \ref{fig:expansion}
shows the position of the molecule cloud along a radial direction as a function of time after the optical trap was abruptly shut off for 1\,ms and then turned back on.  A fit to a damped sine wave (solid curve in
the lower part of Fig. \ref{fig:expansion}) gives a radial
slosh frequency of $174\pm3$ Hz.  To reduce collisional loss of the molecules, these data were taken after removing Rb atoms from the trap \cite{Zirbel2007a}.  For comparison, using the same trapping fields, the measured frequencies for Rb and K atoms are $136\pm2$ Hz and $211\pm4$ Hz, respectively.  The trapping
frequency of the far-off-resonance optical dipole trap depends on the mass
of the particle and its polarizability.  Within the measurement
uncertainties, the ratios of the Rb atom, K atom, and KRb Feshbach
molecule trap frequencies are consistent with the weakly
bound molecule having a mass equal to the sum of the atomic masses
and a polarizability equal to the sum of the atomic
polarizabilities, which is a reasonable assumption for weakly bound molecules.

\section{Binding energy of the Feshbach molecules}

Rf association spectra, such as the one shown in Fig.
\ref{fig:rf_association_spectrum}, can be used to determine the
binding energy of the molecules.  We take the rf frequency at the peak of the molecule association feature and subtract the rf frequency of the atomic Zeeman transition as a measure of the binding energy.  The measured binding energy can have a systematic shift because the initial atom pairs have a range of relative kinetic energies.  We estimate that our measured binding energy could be systematically high by $ 0.6 k_{B} T/h \approx 4$ kHz \cite{Thompson2005}, where $h$ is Planck's constant.  For binding
energies larger than about 0.5 MHz, the molecules are no longer resonant with the imaging light and we are no longer able to see the rf association feature by directly imaging the molecules. However, we
can still detect molecule creation by measuring the loss of K
atoms from the $|9/2, -7/2\rangle$ state. As the magnetic-field detuning from the Feshbach resonance increases, we also find that the maximum conversion fraction decreases, and a longer (as long as 20 s) rf pulse is required.

\begin{figure}
  \centering
  \includegraphics[width=\columnwidth]{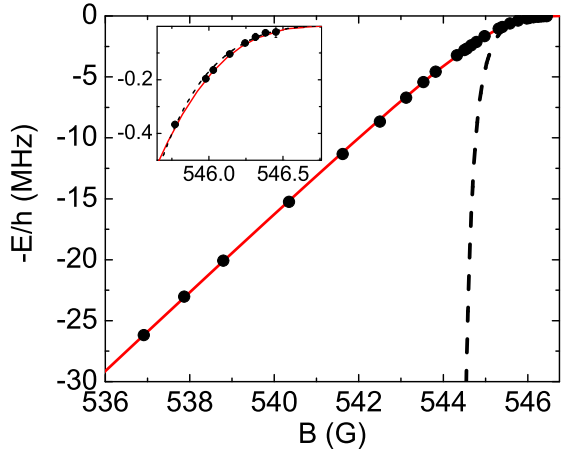}
  \caption{Energy of the heteronuclear Feshbach molecules,
  relative to the Rb $|1,1\rangle$ + K $|9/2,-9/2\rangle$ atomic
  threshold, plotted as a function
  of magnetic field.  The data (solid circles) agree well with our calculation based on a full coupled-channel theory (solid line). Near the Feshbach resonance one expects a universal
  relationship between the s-wave scattering length and the binding energy.  The black, dashed line shows the predicted binding energy using the universal prediction \cite{Gribakin1993}.  Inset: Same as the main plot but
  looking at the region close to the Feshbach resonance.  We fit the inset data to obtain a width of the resonance.}
\label{fig:binding_energy1}
\end{figure}

Figure \ref{fig:binding_energy1} shows the measured molecule binding energy
as a function of the magnetic field.  We plot the negative of the molecule binding energy, $-E/h$, to indicate that the molecule is bound below the atomic Rb $|1,1\rangle$ + K $|9/2,-9/2\rangle$ threshold. We observe
a quadratic dependence of the binding energy at magnetic
fields near the resonance. However, far from the resonance, the
binding energy becomes linear with magnetic field.  The data in Fig.
\ref{fig:binding_energy1} are compared with two theoretical curves. The
dashed curve shows the expected universal behavior very near the
resonance \cite{Gribakin1993,Kohler2003}.  Here, the binding energy is given by $E=\hbar^{2}/(2 \mu_{KRb} (a-\bar{a})^{2})$, where $a$ is the s-wave scattering length, $\mu_{KRb}$ is the $^{40}$K and $^{87}$Rb reduced mass, and $\bar{a}=68.8\,a_{0}$ is the mean scattering length determined by the long-range behavior of the potential. $a_{0}\approx5.29177\times10^{-11}$\,m is the Bohr radius.  Near a Feshbach resonance, the scattering length is given by $a=a_{bg}(1-w/(B-B_{0})) $ \cite{Moerdijk1995, Mies2000}, where $a_{bg}=-185 a_{0}$  for this resonance \cite{Ferlaino2006}.  We determine a width, $w$, by fitting the data from the inset of Fig. \ref{fig:binding_energy1} using the universal binding energy relationship and find $w=-3.6\pm0.1$\,G.  To calculate the molecule binding energy farther from the resonance, we perform a multiple-channel calculation.  The calculated magnetic-field location of the Feshbach resonance does not exactly match with our data.  We therefore fit the experimental data using the coupled-channel calculation and one free parameter, which is the Feshbach resonance position, $B_{0}$.  From this fit, we obtain $B_{0}=546.76\pm0.05$ G.  We find excellent agreement between the data and the
coupled-channel calculation (solid line in Fig. \ref{fig:binding_energy1}).

\begin{figure}
  \centering
  \includegraphics[width=\columnwidth]{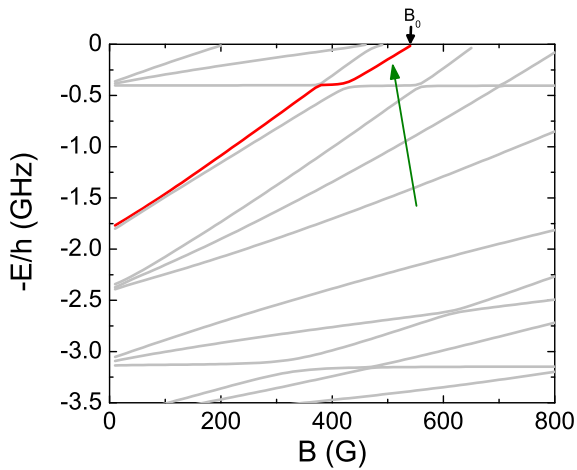}
  \caption{ Calculated molecule energies near the
   K $|9/2,-9/2\rangle$ + Rb $|1,1\rangle$ atomic threshold. All levels
   have total spin projection quantum number $M_{F}=-7/2$.  The
   level energies are calculated using a full coupled-channel
   calculation and have been scaled so that the zero in energy
   is the K $|9/2,-9/2\rangle$ + Rb $|1,1\rangle$ atomic threshold.  The
   bold line is the adiabatic
   level associated with the 546.7 G Feshbach resonance.
} \label{fig:molecule_map}
\end{figure}

To provide a larger view of the Feshbach molecule state that we
probe in the experiment, we show the results of the coupled-channel calculation for a much wider range of magnetic fields in Fig. \ref{fig:molecule_map}.  A Feshbach resonance occurs because of
coupling between an atom scattering state and an energetically
closed bound state.  The channels used in the calculation are
those having a total spin projection quantum number
$M_{F}=m^{Rb}_{f}+m^{K}_{f}=-7/2$ and are therefore coupled to
free atoms in the $|9/2,-9/2\rangle$ and $|1,1\rangle $ states for K
and Rb, respectively.    For the s-wave Feshbach resonance at
$B_{0}=546.76$ G, the dominant closed channel contribution comes from the K
$|7/2,-7/2\rangle$ + Rb $|1,0\rangle$ channel with vibrational quantum
number corresponding to the second level below the dissociation limit in this channel.

We have seen that we can efficiently create heteronuclear Feshbach
molecules starting from an ultracold gas of $^{40}$K and $^{87}$Rb
atoms.  This could provide a starting point for future experimental
efforts aimed at creating polar molecules.  Transfer of the Feshbach
molecules to more deeply bound states could proceed using light or
microwave fields.  Therefore, it is useful to consider how basic
properties of the Feshbach molecules, such as their typical size and
their hyperfine character, depend on the magnetic-field detuning
from the resonance.  We can estimate properties of the molecules
from the measured binding energy curve shown in Fig.
\ref{fig:binding_energy1}.

\section{Properties of the Feshbach molecule}

\begin{figure}
  \centering
  \includegraphics[width=\columnwidth]{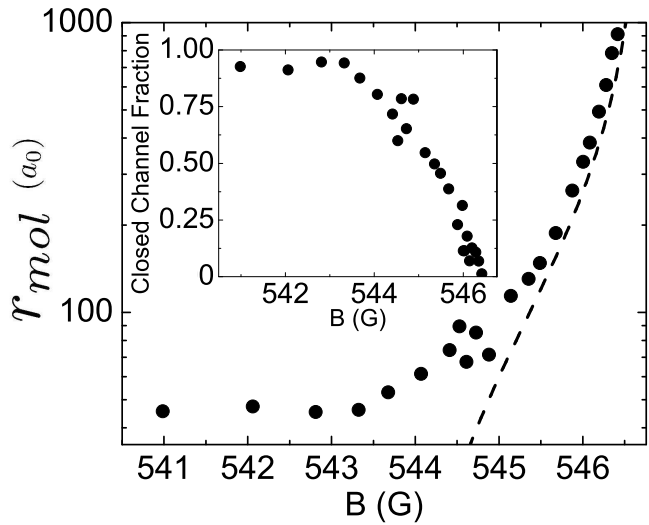}
  \caption{ Estimated molecule size vs. magnetic field near the Feshbach resonance as extracted from the experimentally measured binding energy ($\bullet$). We compare to the universal size prediction, $a/2$ (dashed line).  Inset: Closed channel fraction extracted
  from the experimentally measured binding energy ($\bullet$)(see text).}
\label{fig:binding_energy2}
\end{figure}

The properties of the Feshbach molecule, including its size, are determined by both the open channel and the dominant closed channel.  The bare states associated with these channels have different magnetic moments, and this is, of course, why the molecule's binding energy is magnetic-field tunable. The open and closed
channel contributions to the actual dressed-state molecule can then
be determined from the molecule's magnetic moment relative to the open channel, which is simply the numerical derivative, $\frac{dE}{dB}$, of the measured binding energy shown in Fig. \ref{fig:binding_energy1}. Explicitly, we find the closed channel fraction, $f_c$, from the
following equation: $f_{c}=\frac{1}{\Delta\mu}\frac{dE}{dB}$.  Here, $\Delta\mu=2.38\times \mu_{B}$ is the difference of the atomic magnetic moments of the bare closed and open channels at a magnetic field near the Feshbach resonance and $\mu_{B}$ is the Bohr magneton.   The closed channel
fraction as a function of the magnetic field is shown in the inset of Fig. \ref{fig:binding_energy2}.   For magnetic fields more than 2 G below the Feshbach resonance, we find that $f_{c}>0.5$ indicating that molecule is predominantly closed channel in character.

We can also estimate the molecule size as a function of magnetic
field, $B$.  Near the resonance, where $a$ is large, the molecule size is $ r_{mol}  = a / 2$.   However, far below the resonance, the molecule becomes dominated by the closed channel and the molecule size approaches that of the closed channel molecule. To capture this behavior, we use a simple estimate of the molecule size given by a weighted average of the closed and open channel sizes, $r_{c}$ and $r_{o}$.

\begin{equation}
 r_{mol}= \\
 f_{c} r_{c} + \bigl(1-f_{c}\bigr) r_{o}.
 \label{eq:mol_size}
\end{equation}

$r_{c}$ and $r_{o}$ are calculated using a single-channel model incorporating the correct long-range behavior of the interatomic potential.  Figure \ref{fig:binding_energy2} shows our estimate of the size, $r_{mol}$, as a function of magnetic field near the Feshbach resonance.

To estimate $r_c$ and $r_o$, we numerically solve the Schr\"{o}dinger equation.  Since the loosely bound vibrational wavefunctions have large amplitudes at large internuclear separation, it is important that a model internuclear potential has the correct long-range form.  The short-range details of the potential are not as critical and simply contribute to the overall phase of the wavefunction.  To simplify the calculation, we follow an approximation given by Gribakin and Flambaum \cite{Gribakin1993} and use a model internuclear potential that incorporates the correct long-range behavior and has an impenetrable hard core with a radius $R_{0}$

\begin{equation}
U(r)=
\begin{cases}
\infty & \text{$r < R_0$} \\
-\bigl(\frac{C_{6}}{r^{6}}+\frac{C_{8}}{r^{8}}+\frac{C_{10}}{r^{10}}\bigr) & \text{$r \geq R_0$}.
\end{cases}
\label{eq:potential}
\end{equation}

For $C_{6}$, $C_{8}$, and $C_{10}$ we use the values reported by Pashov \textit{et al.} \cite{Pashov2007}.  $R_{0}$ controls the location of the bound states in the potential and the scattering length.  Since this potential is a single-channel approximation to a multichannel potential, the value of $R_0$ is chosen to be large to avoid the significant divergence between the singlet and triplet potentials that occurs at small internuclear separation. However, $R_{0}$ must be small enough so the potential has an adequate number of bound states.  For $R_{0}\approx 22\,a_{0}$ both criteria are satisfied.

Before calculating the size of the closed channel molecule, we first fine-tune $R_{0}$ so that the calculated background scattering length agrees with the known $^{40}$K $^{87}$Rb s-wave scattering length, $a_{bg}=-185\,a_{0}$.   We note that this procedure yields binding energies of the last three vibrational states that agree to within 7\% with energies obtained using a much more accurate, multichannel potential.  The second-to-last vibrational state with binding energy $E_{-2} \approx h \times 3.2$\,GHz is responsible for the $B=546.7$ G Feshbach resonance. We, therefore, use its wavefunction to calculate the closed-channel size, $r_{c}=\langle r \rangle \approx 40\,a_{0}$.

\begin{figure}
  \centering
  \includegraphics[width=\columnwidth]{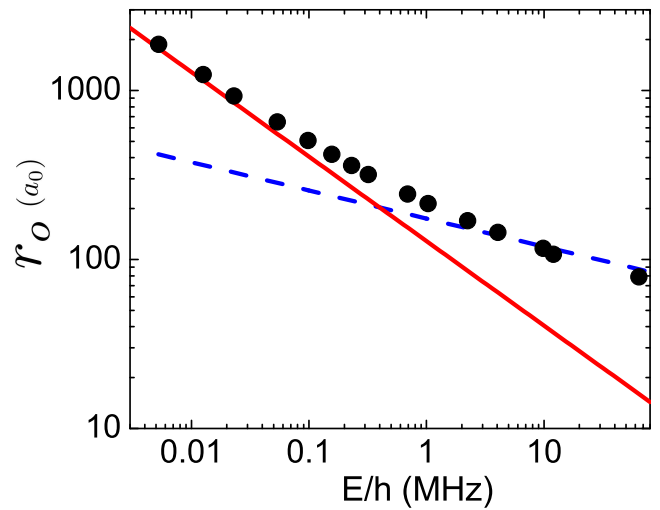}
  \caption{The calculated open-channel molecular size, $r_{o}$, as a function of binding energy.  The solid red line is the universal prediction, $a/2$.  The dashed blue line is the classical outer turning point of the KRb ground state potential. }
\label{fig:ropen}
\end{figure}

Using a similar procedure, we calculate the open-channel size, $r_{o}$, which depends on the magnetic field detuning from the Feshbach resonance.  In this case, however, we vary $R_0$ to control the binding energy, $E$, of the weakest bound molecular state.  For each $E$, we calculate the size, $r_{o}=\langle r \rangle$, from the calculated molecular wavefunction.  For the smallest binding energies, the calculated size agrees with the universal prediction.  At larger binding energies the molecular size is approximately the classical turning point.  Figure \ref{fig:ropen} shows the behavior of $r_{o}$.

\section{Conclusions}

In conclusion, we have used rf association to create heteronuclear
Feshbach molecules from an ultracold gas of $^{40}$K and $^{87}$Rb
atoms confined in an optical dipole trap.  Unlike previous work
\cite{Ospelkaus2006}, we do not isolate pairs of atoms in individual wells of a deep optical lattice potential.  Consequently, the atoms and the
molecules are free to collide and pairs of atoms are initially in a near-continuum of motional quantum states. Nevertheless, by
optimizing the pulse duration and frequency of the rf, as well as
the initial atom gas temperature, we achieve comparable overall
efficiency in converting atoms to weakly bound Feshbach molecules. We also
demonstrate that these molecules are ultracold and trapped.

A gas of ultracold, heteronuclear Feshbach molecules can provide a
starting point for optical manipulation to create a sample of
ultracold polar molecules.  In anticipation of such experiments, we
have measured the magnetic-field dependent binding energy of the
Feshbach molecules.  We extend these measurements to magnetic fields
much farther from resonance than what is typically of interest for
ultracold atom experiments.  We measure binding energies as large as
26 MHz and find that for magnetic fields more than 2 or 3 G detuned
from resonance, the binding energy curve is linear. This means that
by lowering the magnetic field to a few gauss below the resonance,
one can have Feshbach molecules that are predominantly closed
channel in character with the corresponding molecular size.

\begin{acknowledgments}
We thank B. Neyenhuis, A. Wilson, and C. Ospelkaus for experimental
assistance, A. Pe'er for useful discussions; K.-K. Ni acknowledges support from the NSF and S. O. from the A.-v.Humboldt Foundation.  We acknowledge funding support from NIST, NSF, and the W. M. Keck Foundation.
\end{acknowledgments}


\end{document}